\documentclass[twocolumn,superscriptaddress,dvipsnames,longbibliography]{revtex4}
\usepackage{graphicx,xcolor,array,hhline}
\usepackage{bbold,mathtools,amsmath,amssymb,amsthm,amsfonts}
\usepackage{mathptmx}
\usepackage{hyperref}
\usepackage{xfrac}
\usepackage{url}
\usepackage{subfigure}

\usepackage{algorithm} 
\usepackage{algpseudocode,pifont}

\floatname{algorithm}{Protocol}
\usepackage[labelsep=period]{caption}
\usepackage{ftnxtra}

\theoremstyle{plain}

\theoremstyle{definition}

\newcommand{\bra}[1]{\langle#1|}
\newcommand{\ket}[1]{|#1\rangle}

\def\>{\rangle} 
\def\<{\langle}

\begin{document}
\title{Computing on quantum shared secrets}

\author{Yingkai Ouyang}\email{yingkai\_ouyang@sutd.edu.sg}
\affiliation{Singapore University of Technology and Design, 8 Somapah Road, Singapore 487372}
\author{Si-Hui Tan}
\affiliation{Singapore University of Technology and Design, 8 Somapah Road, Singapore 487372}
\author{Liming Zhao}
\affiliation{Singapore University of Technology and Design, 8 Somapah Road, Singapore 487372}
\author{Joseph F. Fitzsimons}
\affiliation{Singapore University of Technology and Design, 8 Somapah Road, Singapore 487372}
\affiliation{Centre for Quantum Technologies, National University of Singapore, 3 Science Drive 2, Singapore 117543}

\begin{abstract}
A ($k$,$n$)-threshold secret-sharing scheme allows for a string to be split into $n$ shares in such a way that any subset of at least $k$ shares suffices to recover the secret string, but such that any subset of at most $k-1$ shares contains no information about the secret. Quantum secret-sharing schemes extend this idea to the sharing of quantum states. Here we propose a method of performing computation on quantum shared secrets. We introduce a ($n$,$n$)-quantum secret sharing scheme together with a set of protocols that allow quantum circuits to be evaluated on the shared secret without the need to decode the secret. We consider a multipartite setting, with each participant holding a share of the secret. We show that if there exists at least one honest participant, no group of dishonest participants can recover any information about the shared secret, independent of their deviations from the protocol.
\end{abstract}

\maketitle

The connected nature of modern computing infrastructure has led to the widespread adoption of distributed and delegated computation \cite{armbrust2010view}, with hard computational tasks routinely delegated to remote computers. In such a setting, security of the computation can be a very real concern. For several decades it has been understood that quantum cryptography offers stronger security for key distribution than is possible using purely classical communication over untrusted channels \cite{BB84,Eke91}. More recently, quantum protocols have appeared for secure computation tasks such as secure multi-party computation \cite{CGS02}, blind computation \cite{BFK09,morimae2012blind,barz2012demonstration,broadbent2015delegating} and verifiable delegated computation \cite{ABE,RUV,fitzsimons2012unconditionally,morimae2014verification,hayashi2015verifiable}. In the present manuscript, we focus on a different form of secure computation, namely the evaluation of quantum circuits on shared secrets.

In a secret sharing scheme, an $r$-bit string {\bf r} which is meant to be kept a secret, is encrypted into an $s$-bit string {\bf s}. These $s$ bits are subsequently distributed among $n$ parties, with the intention that whenever too few of the parties collude, the colluding parties cannot perfectly recover the secret message {\bf r}. The reversibility of the encryption allows the secret message {\bf r} to be recovered when all of the $n$-parties assemble the data that they were distributed. In a $(k,n)$-threshold scheme for classical secret sharing \cite{Shamir1979,Blakley1979}, it is required that no group with fewer than $k$ colluding parties can reconstruct the secret message {\bf r}, and any $k$ parties can reconstruct {\bf r}. Similarly in a $(k,n)$-threshold quantum secret sharing scheme, a secret quantum state of $s$ qubits is shared among $n$ parties such that no group fewer than $k$ colluding parties can reconstruct the secret quantum state \cite{HBB99,CGL99,Got00,zhang2005multiparty,DSa08}, and any $k$ parties can reconstruct the secret quantum state. Here, we present an $(n,n)$-threshold quantum secret sharing scheme that also supports evaluation of quantum circuits on the shared secret. 

Our secret sharing scheme with computation can be seen as a form of secure delegated multipartite quantum computation where the delegated computation is made public. We emphasize that our scheme is not naturally a blind quantum computation scheme, because blind quantum computation also requires the intended quantum computation kept secret from the evaluator \footnote{However, by fixing the public function to act as a programmable computer, it is possible to implement a form of blind computation, similar to other secure computing protocols \cite{dunjko2014composable}.}. As such, the setting we consider is more closely related to that of quantum homomorphic encryption schemes \cite{Rohde,tan2016quantum,ouyang2015quantum,BJe15,DSS16}, 
which allows the quantum computation to be performed to be public 
and requires the decoding algorithm to be independent of the depth of the computation. Indeed, we are motivated by a particular quantum homomorphic encryption scheme, introduced in \cite{ouyang2015quantum}, that supports transversal evaluations of Clifford gates, and present a secret sharing scheme that allows the evaluation of Clifford gates by requiring the $n$ non-interacting parties to perform the corresponding Clifford operations in parallel. A constant number $t$ of non-Clifford gates can also be implemented via a coordinated gate teleportation using logical magic states. Since the security of our scheme is independent of the security of the quantum homomorphic encryption scheme in Ref.~\cite{ouyang2015quantum}, the no-go results for fully quantum homomorphic encryption schemes with both perfect \cite{YPF14} and imperfect \cite{NewmanShi2016} information theoretic security do not limit the class of circuits which can be evaluated.

\begin{figure}[t]\centering
        \includegraphics[width=0.7\columnwidth]{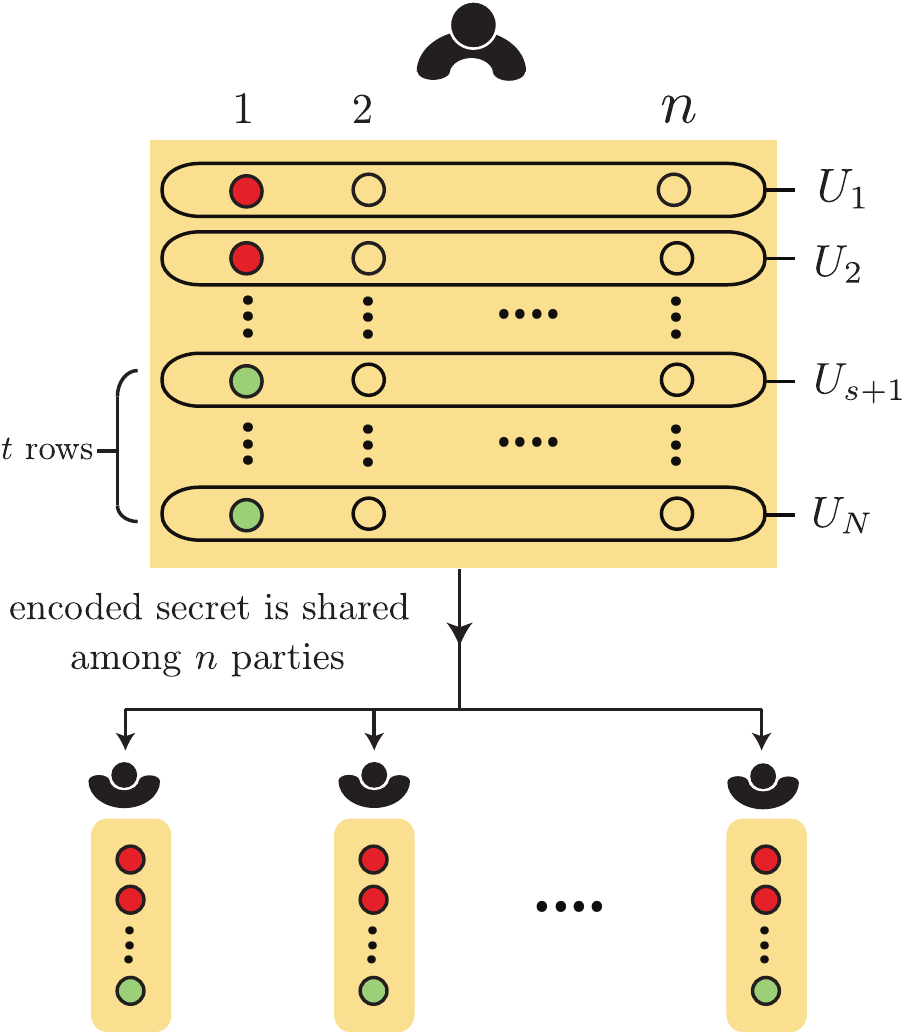}
	\caption{The upper portion of the figure shows the secret, which is an $s$-qubit state located on the first column. The qubits carrying the secret are shaded in red. The remaining $t$ states located on the first column are each initialized as a magic state. These qubits are shaded in green. The unshaded qubits are initialized in the maximally mixed state. The unitaries $U_1,\dots, U_N$ spread the states from qubits in the first column to qubits in the remaining columns, such that the encoded secret resides in the first $s$ rows of qubits. Each party receives a single column of the encoded secret, together with their part of the logical magic states.  \label{fig:dataqubit}
	} 
\end{figure}

Our secret sharing scheme comprises of four procedures as described in Protocol \ref{prot:sharing}. We label qubits according to a 2-dimensional arrangement as depicted in Fig.~\ref{fig:dataqubit}. In the input procedure of Protocol \ref{prot:sharing}, $N = s+t$ qubits are initialized on a single column, with the first $s$ qubits containing the quantum secret, and the last $t$ qubits each initialized in the magic state $\tau = \frac{I}{2} + \frac{X+Y}{2 \sqrt 2}$, where $I$, $X$, $Y$, and $Z$ are the usual Pauli matrices. These magic states are consumed during the evaluation in reverse order, starting from the last row. 
We focus on the case where $n-1$ is divisible by 4. This is not a limiting factor, since if this is not the case, one can simply prepare $\left\lceil \frac{n-1}{4} \right\rceil + 1$ shares and give multiple shares to a single party. In the encoding procedure of Protocol \ref{prot:sharing}, $n-1$ additional columns of $N$ qubits in the maximally mixed state are appended. This yields an $Nn$-qubit quantum state arranged in a grid with $N$ rows and $n$ columns. Subsequently a unitary encoding $U$ is applied on the $Nn$ qubits, which spreads the quantum secret from the first column to all the $n$ columns. Here $U = U_1 \otimes \dots \otimes U_N$ is a tensor product of the unitaries $U_1,\dots, U_N$, where each $U_x$ acts only on the $x$-th row of qubits and comprises of only CNOT gates. Specifically $U_x = B_x A_x$, where (i) $A_x$ comprises of $n-1$ commuting CNOT gates with controls all on the first column and targets on each of the remaining columns, and (ii) $B_x$ comprises of $n-1$ commuting CNOT with targets all on the first column and controls on every other column. Although $U_x$ is a fixed unitary, the induced encoding is random because $n-1$ of the qubits that $U_x$ acts on are random; the qubits from the second column to the last column are initialized as either $|0\>$ or $|1\>$ with probability $\sfrac 1 2$. This random encoding maps the quantum secret into a random code which is a highly mixed state, similar to Ref.~\cite{ouyang2015quantum}. In the sharing procedure of Protocol \ref{prot:sharing}, the $Nn$-qubit quantum state is shared equally among $n$ parties, with each party receiving a single column of $N$ qubits. In decoding procedure of Protocol \ref{prot:sharing}, the $n$ shares are assembled, the inverse encoding circuit $U ^\dagger$ is performed, and all but the first column of qubits are discarded, which yields the $N$-quantum secret on the first column.
 
\begin{algorithm}[t]
\caption{Secret sharing scheme \label{prot:sharing}}
Here, $\mathcal H_{x,y}$ labels the qubit on the $x$-th row and the $y$-th column, 
and $\mathcal R_{x}$ labels the qubits on the $x$-th row.

\begin{enumerate}
	\item \textbf{Input:} \label{prot:sharing:input} From the $s$-qubit quantum secret, 
		assign the $x$-th qubit to $\mathcal H_{x,1}$ for $x=1,\dots ,s$.
		Assign $\tau$ to each of $\mathcal H_{N-k+1,1}, \dots, \mathcal H_{N,1}$.
  	\label{prot:sharing:input}
	
	\item \textbf{Encoding:} \label{prot:sharing:encoding}
		To prepare the $x$-th logical qubit for $x=1, \dots, N$:
		\begin{enumerate}
			\item Prepare each of $\mathcal H_{x,2}, \dots, \mathcal H_{x,n}$ in state $\frac{I}{2}$. \label{step:init}
				\item Apply $A_x$: Perform a CNOT with control on $\mathcal H_{x,1}$ 
					and target on $\mathcal H_{x,y}$ for every $y = 2, \dots, n$.
				\item Apply $B_x$: Perform a CNOT with target on $\mathcal H_{x,1}$ 
					and control on $\mathcal H_{x,y}$ for every $y = 2, \dots, n$.
		\end{enumerate}

	\item \textbf{Sharing:} Assign the qubits in the $y$-th column to the $y$-th share for $y=1,\dots, n$.  \label{prot:sharing:sharing}
 
	\item \textbf{Decoding:}\label{prot:sharing:decoding}
		\begin{enumerate}
			\item Assemble the $n$ shares. 
			\item For each $x = 1, \dots, N$, implement $B_x$ followed by $A_x$ on $\mathcal R_x$.
			\item Output the qubits in the first column, discarding all other qubits.
		\end{enumerate}

\end{enumerate}

\end{algorithm} 
 
 %%%%%%%%%%%%% figure %%%%%%%%%%%%%%%%%%%%%%%%%%%%%%%%%%%
%%%%%%%%%%%%%%%%%%%%%%%%%%%%%%%%%%%%%%%%%%%%%%%%%%%%%%%%%

To evaluate a quantum circuit on the shared secret, each party performs quantum computation only on their share of the quantum state. We consider the approximately universal model of quantum computation based on a discrete set of gates composed of Clifford group gates and a single non-Clifford group gate, in this case $T = |0\>\<0| +  e^{i \pi/4}|1\>\<1|$ although other choices are possible. As we shall see, quantum circuits composed of arbitrarily many Clifford gates and up to some constant number $t$ of $T$-gates can be evaluated on the shared secret. We will consider the evaluation of a sequence $V=( V_1,\dots, V_L)$ of such gates on the $s$-qubit quantum secret shared by $n$ parties. The gates $V_1,\dots, V_L$ are unitary matrices on $s$ qubits and are assumed to be known to every party. Using the knowledge of $V$, each party implements a sequence of operations on their share of the qubits, as specified in Protocol \ref{prot:compute}. The computation is performed after the sharing procedure and before the decoding procedure of Protocol \ref{prot:sharing}, as we now describe.

\begin{algorithm} 
\caption{Gate evaluation on shared quantum secret}\label{prot:compute}
Given a gate $V_i$ to be evaluated on the shared secret:
\begin{itemize}
\item \textbf{Clifford group:} If $V_i$ is in the Clifford group each party applies $V_i$ to their share.
\item \textbf{$T$-gates:} If $V_i$ is a $T$-gate	 on qubit $j$, each party $y$ does as follows
\begin{enumerate}
\item Apply a CNOT gate controlled by qubit $j$ and targeted on qubit $N-k+1$.
\item Apply a CNOT gate controlled by qubit $N-k+1$ and targeted on qubit $j$.
\item Measure qubit $N-k+1$ in the computational basis, and broadcast the result $m_y$.
\item If the parity of $\mathbf{m} =  (m_1 ,\ldots, m_n)$ is odd, apply the correction operator $SX$ to qubit $j$.
\end{enumerate}
\end{itemize}
\end{algorithm}

\begin{figure}[t]\centering
        \includegraphics[width=\columnwidth]{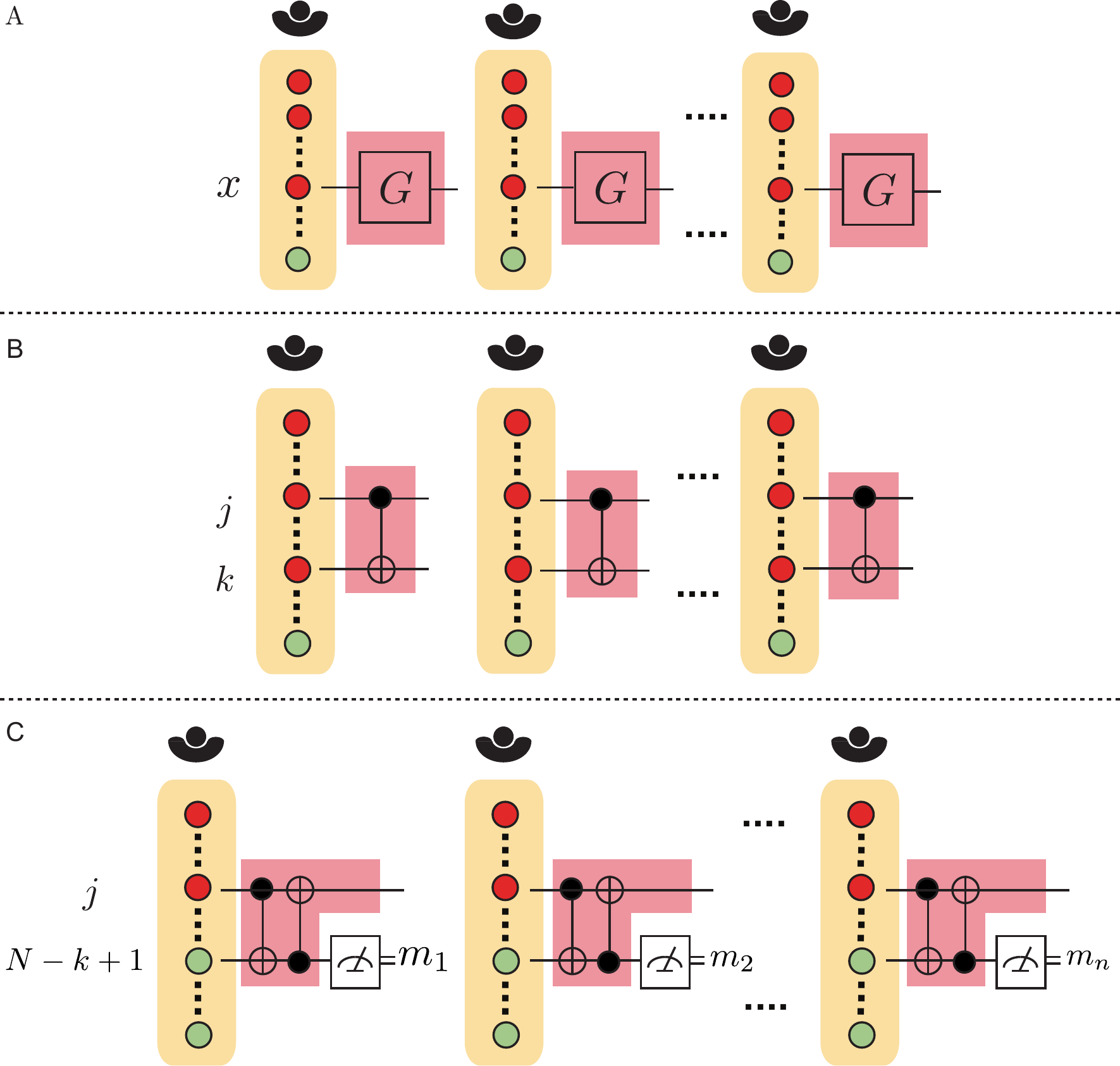}
	\caption{In this figure, all qubits that are part of the secret are shaded in red, while those that are part of the logical magic states are shaded in green. A) Multipartite implementation of a logical single-qubit Clifford gate $G$ on the $x$-th row. B) Multipartite implementation of a logical CNOT operator with the control qubit and the target qubit on the $j$-th and $k$-th last row respectively. C) A logical gate teleportation protocol that implements a logical $T$-gate on the $j$-th logical qubit without the measurement-dependent Clifford correction is depicted. Collectively, the qubits on the row that is later measured are initialized in logical magic state. To implement the correction, the measurement outcomes are made public, and each party applies 
a single Clifford gate $SX$ on the $j$-th qubit only when the parity of $\bf m$ is odd.	 \label{fig:transversal}} 
\end{figure}

When $V_i$ is a Clifford gate that applies non-trivially on some set of logical qubits, each party performs $V_i$ on the corresponding subset of their column of qubits, thereby collectively implementing $V_i^{\otimes n}$. This procedure is depicted in Fig.~\ref{fig:transversal}A for single qubit Clifford gates, and Fig.~\ref{fig:transversal}B for a CNOT gate. Let $\mathcal P = \{I, X,Y,Z\}$ denote the set of the Pauli matrices. Then the fact that $n-1$ is divisible by 4 implies that for $\sigma \in \mathcal P$, 
\begin{align}
U_x  (\sigma \otimes {I} ^{\otimes n-1}) U_x^\dagger = \sigma ^{\otimes n}. \label{eq:transversal-Cliffords}
\end{align}
Since $V_i$ is in the Clifford group, it maps the Pauli group onto itself, 
\begin{align}
U_x  (V_i \sigma V_i^\dagger \otimes {I} ^{\otimes n-1}) U_x^\dagger = V_i^{\otimes n} \sigma ^{\otimes n} \left(V_i^\dagger\right)^{\otimes n}.	
\end{align} 
Hence the transversal Clifford group gates correspond to the logical Clifford group gates on our random codespace \cite{ouyang2015quantum}. 

It is also possible to perform a constant number $t$ of $T$-gates on the quantum secret via gate teleportation. For each $T$-gate that is to be performed, a logical magic state 
\begin{align}\label{eq:tautilde}
{\widetilde \tau} = \frac{I^{\otimes n}}{2^n}+\frac{X^{\otimes n}+Y^{\otimes n}}{2^n\sqrt{2}}
\end{align}
must be prepared. This is achieved by the input and encoding procedures of Protocol \ref{prot:sharing}, however we do not rule out the possibility of replacing this pre-sharing of magic states with a procedure for the parties to interactively prepare states on demand without the involvement of the initial sharer. Each of these logical magic states is taken to be located on the last $t$ rows of the encoding. To prepare ${\widetilde \tau}$ on the $x$-th row, the first qubit in the $x$-th row is initialized as $TH|0\>$ with the remaining qubits prepared in the maximally mixed state. 
The encoding unitary $U_x$ is subsequently applied. 
To evaluate the $k$-th $T$-gate on qubit $j$ of the shared secret, each party proceeds as follows. They first apply a CNOT with control on the $j$-th qubit and target on the $k$-th last qubit of their share. They then apply a CNOT with control on the $k$-th last qubit and target on the $j$-th qubit. Each party $y$ then measures the $k$-th last qubit in the $\{|0\>,|1\>\}$ basis and broadcasts the measurement result $m_y$ to every other party over a public classical channel. Lastly, if the parity of the measurement results $\bf m$ is odd, each party applies a single-qubit Clifford gate $SX$ on the $j$-th qubit. If the parity is even, no such correction is necessary. This procedure is depicted in Fig.~\ref{fig:transversal}C. The evaluation of each $T$-gate in this way amounts to the logical implementation of a gate teleportation protocol that consumes one magic state \cite{ZLC00}. 

We now describe how the evaluation of the $T$-gate works by explicitly considering the operations that the $n$ parties implement. Denoting $\overline I =I ^{\otimes n}$, $\overline X =X ^{\otimes n}$, $\overline Y =Y ^{\otimes n}$ and $\overline Z =Z ^{\otimes n}$, the correct implementation of a logical $T$-gate on the state $\widetilde{\rho}=2^{-n}(\overline I+a\overline X+b \overline Y+c \overline Z)$ 
shared by the $j$-th qubit of each party must yield
\begin{align}\frac{1}{2^{n}} \left(\overline I
+\frac{\left(a-b\right)}{\sqrt{2}}\overline X
+\frac{\left(a+b\right)}{\sqrt{2}}\overline Y
+c\overline Z\right).
\label{eq:T-correctness} 
\end{align}  
This follows from the conjugation relations for the $T$-gate given by $TXT ^\dagger=\frac{1}{\sqrt{2}} (X+Y)$, $TYT ^\dagger =\frac{Y-X}{\sqrt 2}$, and $TZT ^\dagger =Z$. 
After every party applies the CNOT gates as depeicted in Fig.~\ref{fig:transversal}C, 
the joint quantum state on the $j$-th qubit and the $k$-th last qubit of every party given by  $\widetilde{\rho} \otimes {\widetilde \tau}$
is mapped to the state
\begin{align} 
&\frac{ 
       \overline I \otimes \overline I 
+ a \overline I  \otimes \overline X 
+ b \overline Z \otimes \overline Y
+ c \overline Z \otimes \overline Z
}{2^{2n}}  \notag\\
+&
\frac{ 
\overline X   \otimes \overline X 
+ a \overline X  \otimes \overline I 
+ b  \overline Y \otimes \overline Z
- c  \overline Y \otimes  \overline Y
}{2^{2n} \sqrt 2}  \notag\\
+&
\frac{  
\overline Y \otimes \overline X  
+ a  \overline Y \otimes \overline I
- b  \overline X \otimes  \overline Z  
+ c \overline X  \otimes  \overline Y
}{2^{2n} \sqrt 2} . \label{eq:teleport-CNOT}
\end{align}
To show this, we have used the commutation relations of the CNOT with various two-qubit Pauli matrices \cite[Eqs.~(4.32)-(4.37)]{nielsen-chuang}. The parity of $\bf m$ is equivalent to the observable $\overline Z$ on the $k$-th last qubit of each share. If the parity is even,
the resultant state on the $j$-th qubit of every party is collectively
\begin{align}
\widetilde{\rho}_{even} 
&=
\frac{\overline I }{2^n}
+\frac{\left(a-b\right) \overline  X }{2^n\sqrt{2}}
+\frac{\left(a+b\right) \overline Y }{2^n\sqrt{2}}
+\frac{c \overline Z }{2^n},
\end{align}
and the evaluation of the $T$-gate is successful. If the parity is odd, however, the resultant state of these qubits is
\begin{align}
\widetilde{\rho}_{odd} 
&=
\frac{\overline I }{2^n}
+\frac{\left(a+b\right) \overline X }{2^n\sqrt{2}}
+\frac{\left(a-b\right) \overline Y }{2^n\sqrt{2}}
-\frac{c \overline Z }{2^n}.
\end{align}
Applying $SX$ to each qubit transforms the state into $\widetilde{\rho}_{even}$, resulting in a correct evaluation of the $T$-gate.

We now turn to the issue of security. This requires several steps. 
First we show that the scheme outlined in Protocol \ref{prot:sharing} is a $(n,n)$-threshold secret sharing scheme as claimed. We then prove that the evaluation of quantum circuits on the shared secret performed according to Protocol \ref{prot:compute} does not compromise the encoding. Specifically, we show that no subset of $n-1$ parties can collude to produce any state correlated with the encoded secret, beyond any prior knowledge they may have.

A $(k,n)$-threshold quantum secret-sharing scheme \cite{CGL99,Got00} is a quantum operation that maps a secret quantum density matrix to an encoded state that can be divided among $n$ parties such that (1) any $k$ or more parties can perfectly reconstruct the secret quantum state, and (2) any $k-1$ or fewer parties can collectively deduce no information at all about the secret quantum state. 
The first property is trivially satisfied by Protocol \ref{prot:sharing} when $k=n$, since the encoding procedure is perfectly reversible with inverse operation given by the specified decoding procedure. Turning to the second property, we consider the result of encoding a state 
\begin{equation}
	\rho_\text{secret} = 2^{-s} \sum_{ \sigma \in \mathcal P^{\otimes s}} w_{\sigma} \sigma
\end{equation}
according to Protocol \ref{prot:sharing}. Here $\sigma=\sigma_1 \otimes \ldots \otimes \sigma_{s}$ and $w_\sigma = 1$ when $\sigma$ is the trivial Pauli operator, $\sigma=I ^{\otimes s}$. It is the coefficients $w_\sigma$ for the non-trivial Pauli operators $\sigma$ in $\mathcal P ^{\otimes s}$	that collectively define the quantum secret. From Eq.~\ref{eq:transversal-Cliffords} it follows that the resulting state is
\begin{equation}
	\tilde{\rho}_\text{secret} = 2^{-s} \left(\sum_{ \sigma \in \mathcal P^{\otimes s}} w_{\sigma} \sigma^{\otimes n} \right) \otimes \tilde{\tau}^{\otimes t},\label{eq:sharedstate}
\end{equation}
where the tensor product in $\sigma^{\otimes n}$ is taken across different shares of the secret. Property (2) then follows trivially, since the reduced density matrix for any subsystem of $n-1$ shares (i.e. $n-1$ columns) is necessarily the maximally mixed state, because all non-trivial $\sigma$ are traceless. 

Turning to the issue of the security of Protocol \ref{prot:compute}, we consider the state of the system across a bipartition between a single honest party, who follows the protocol, and the remaining $n-1$ parties who are not restricted in their actions. We now show that the bits broadcast by the honest party are uniformly random and independent of the actions of the other parties. Given a sequence of gates $(V_1,\ldots, V_L)$ with the honest party acting as described by Protocol \ref{prot:compute}, our strategy is to show that after evaluation of the $\ell$-th gate, the state of the system has the form
\begin{align} 
 \rho^{(\ell)}_{\rm joint}=\sum_{ \substack{
 \sigma  \in \mathcal P^{\otimes s} \\
 \theta \in \{I,X,Y\}^{\otimes t-k}
 }}  
 b_{\sigma,\theta}^{(\ell)}
\left(\frac{\sigma \otimes \theta}{2^N}\right)
			\otimes \chi_{\sigma,\theta}^{(\ell)}, \label{eq:rhostructure}
 \end{align} 
where $k\leq \ell$ is the number of $T$-gates in $(V_1,\ldots, V_\ell)$, $\{b_{\sigma,\theta}^{(\ell)}\}$ is a set of scalars, and $\{\chi_{\sigma,\theta}^{(\ell)}\}$ is a set of operators on the Hilbert space representing the system of the dishonest parties. 
We have excluded the honest party's measured qubits, as these are in a product state with the rest of the system.

The proof proceeds by induction. 
We assume that the system is in a state $\rho^{(\ell-1)}_{\rm joint}$ of the form of Eq.~\ref{eq:rhostructure} after evaluation of the first $\ell - 1$ gates. 
If $V_\ell$ is a Clifford group gate, then the honest party applies $V_\ell$ on some subset of the first $s$ qubits of their share, 
while the dishonest parties are free to perform any completely positive and trace preserving map on their side of the bipartition. 
Since $V_\ell I^{\otimes s} V_\ell^\dagger = I^{\otimes s}$ and $V_\ell \mathcal{P}^{\otimes s} V_\ell^\dagger = \mathcal{P}^{\otimes s}$, and since the operation applied by the dishonest parties on their side of the bipartition is linear, the resulting state $\rho^{(\ell)}_{\rm joint}$ is in the form of Eq.~\ref{eq:rhostructure} as claimed. 
When $V_\ell$ is a $T$-gate on some qubit $j$, the situation is more complicated. 
Since the actions of the honest party only affect the $j$-th qubit and $k$-th last qubit of their share, we will consider the effect of these actions on all combinations of Pauli operators on these two qubits which can have non-zero coefficients in $\rho^{(\ell-1)}_{\rm joint}$. From the induction assumption, these are given by the first column of Table \ref{table:Tgate}. The effect of the honest party applying CNOT operations as prescribed by the first two steps of the $T$-gate procedure in Protocol \ref{prot:compute} is to transform these operators into the corresponding Pauli operators given by the second column of Table \ref{table:Tgate}. 
As the operator $I\otimes Z$ does not appear, it follows that the expectation for $m_\text{H}$, the measurement result of the honest party's measurement, is precisely zero. Hence $m_\text{H}$ is uniformly random and independent of the non-trivial weights $\{b_{\sigma,\theta}\}$. The effect of the measurement on the Pauli operators is given by the third column of Table \ref{table:Tgate}, 
which implies that the resulting state is in the form of Eq.~\ref{eq:rhostructure}. 
Since the correction $SX$ is merely a local Clifford group operator, the final state $\rho_\text{joint}^{(\ell)}$ is always of the correct form independent of the parity of $\bf m$. Since the initial state after sharing, given by Eq.~\ref{eq:sharedstate} is of the form of Eq.~\ref{eq:rhostructure}, the induction hypothesis holds for all $0 \le \ell \le L$, and the measurement results of the honest party convey no information which can be used by the dishonest participants to recover $\rho_\text{secret}$.

\begin{table}[t]
\begin{tabular}{| >{\centering\arraybackslash}m{2cm} | >{\centering\arraybackslash}m{2cm} | >{\centering\arraybackslash}m{4cm} |} 
 \hline 
\vspace{0.8mm} $\sigma_j\otimes \theta_k$ \vspace{0.8mm} & $\tau_{j,k}$ & $(I\otimes \bra{m_\text{H}})\tau_{j,k}(I\otimes \ket{m_\text{H}})$ \\ \hhline{|=|=|=|}
  $I\otimes I$   & $I\otimes I$  & $I$ \\
 $I\otimes X$ &   $X\otimes X$  & $0$\\
 $I\otimes Y$ & $Y\otimes X$ & $0$\\
$X\otimes I$ & $I\otimes X$ & $0$\\
$X\otimes X$ & $X\otimes I$ & $X$\\
$X\otimes Y$ & $Y\otimes I$ & $Y$\\
$Y\otimes I$ & $Z\otimes Y$ & $0$\\
$Y\otimes X$ & $Y\otimes Z$ & $(-1)^{m_\text{H}}Y$\\
$Y\otimes Y$ & $-X\otimes Z$ & $(-1)^{m_\text{H}+1}X$\\
$Z\otimes I$ & $Z\otimes Z$ & $(-1)^{m_\text{H}}Z$\\
$Z\otimes X$ & $Y\otimes Y$ & $0$\\
$Z\otimes Y$ & $X\otimes Y$ & $0$\\
 \hline
\end{tabular}
\caption{Table of values of (i) $\sigma_j \otimes \theta_k$, (ii) the resulting operator $\tau_{j,k}$ after the application of steps 1 and 2 of the $T$-gate procedure of Protocol \ref{prot:compute}, and (iii) $(I\otimes \bra{m_\text{H}})\tau_{j,k}(I\otimes \ket{m_\text{H}})$ for $\sigma_k\in \mathcal P$, $\theta_k\in\{I, X, Y\}$.}\label{table:Tgate}
\end{table}  
 
The scheme we have presented above, therefore, represents a $(n,n)$-threshold secret sharing scheme that also allows for the evaluation of quantum circuits on the shared secret without lowering the threshold. While the complexity of such circuits is limited in terms of the number of $T$-gates to the number of corresponding magic states incorporated in the initial sharing, whether it is possible to create such states as needed without involvement of the initial sharer presents an interesting avenue for future research. Intuitively, the security of our scheme
is based on a randomized error correction code which leaves only weight $n$ operators constant while admitting transversal Clifford gates. This suggests that the use of less random error-correction codes will allow for $(k,n)$-threshold schemes for other values of $k$.

The authors thank Mahboobeh Houshmand and Monireh Houshmand for useful discussions.
The authors acknowledge support from Singapore's Ministry of Education and National Research Foundation. JFF and SHT acknowledge support from the Air Force Office of Scientific Research under AOARD grant FA2386-15-1-4082. This material is based on research funded in part by the Singapore National Research Foundation under NRF Award NRF-NRFF2013-01.

\bibliographystyle{apsrev}
\bibliography{ssc}

\begin{thebibliography}{30}
\expandafter\ifx\csname natexlab\endcsname\relax\def\natexlab#1{#1}\fi
\expandafter\ifx\csname bibnamefont\endcsname\relax
  \def\bibnamefont#1{#1}\fi
\expandafter\ifx\csname bibfnamefont\endcsname\relax
  \def\bibfnamefont#1{#1}\fi
\expandafter\ifx\csname citenamefont\endcsname\relax
  \def\citenamefont#1{#1}\fi
\expandafter\ifx\csname url\endcsname\relax
  \def\url#1{\texttt{#1}}\fi
\expandafter\ifx\csname urlprefix\endcsname\relax\def\urlprefix{URL }\fi
\providecommand{\bibinfo}[2]{#2}
\providecommand{\eprint}[2][]{\url{#2}}

\bibitem[{\citenamefont{Armbrust et~al.}(2010)\citenamefont{Armbrust, Fox,
  Griffith, Joseph, Katz, Konwinski, Lee, Patterson, Rabkin, Stoica
  et~al.}}]{armbrust2010view}
\bibinfo{author}{\bibfnamefont{M.}~\bibnamefont{Armbrust}},
  \bibinfo{author}{\bibfnamefont{A.}~\bibnamefont{Fox}},
  \bibinfo{author}{\bibfnamefont{R.}~\bibnamefont{Griffith}},
  \bibinfo{author}{\bibfnamefont{A.~D.} \bibnamefont{Joseph}},
  \bibinfo{author}{\bibfnamefont{R.}~\bibnamefont{Katz}},
  \bibinfo{author}{\bibfnamefont{A.}~\bibnamefont{Konwinski}},
  \bibinfo{author}{\bibfnamefont{G.}~\bibnamefont{Lee}},
  \bibinfo{author}{\bibfnamefont{D.}~\bibnamefont{Patterson}},
  \bibinfo{author}{\bibfnamefont{A.}~\bibnamefont{Rabkin}},
  \bibinfo{author}{\bibfnamefont{I.}~\bibnamefont{Stoica}},
  \bibnamefont{et~al.}, \bibinfo{journal}{Communications of the ACM}
  \textbf{\bibinfo{volume}{53}}, \bibinfo{pages}{50} (\bibinfo{year}{2010}).

\bibitem[{\citenamefont{Bennett and Brassard}(1984)}]{BB84}
\bibinfo{author}{\bibfnamefont{C.~H.} \bibnamefont{Bennett}} \bibnamefont{and}
  \bibinfo{author}{\bibfnamefont{G.}~\bibnamefont{Brassard}}, in
  \emph{\bibinfo{booktitle}{Proceedings of IEEE International Conference on
  Computers, Systems and Signal Processing}} (\bibinfo{organization}{New York},
  \bibinfo{year}{1984}), vol. \bibinfo{volume}{175}.

\bibitem[{\citenamefont{Ekert}(1991)}]{Eke91}
\bibinfo{author}{\bibfnamefont{A.~K.} \bibnamefont{Ekert}},
  \bibinfo{journal}{Phys. Rev. Lett.} \textbf{\bibinfo{volume}{67}},
  \bibinfo{pages}{661} (\bibinfo{year}{1991}),
  \urlprefix\url{http://link.aps.org/doi/10.1103/PhysRevLett.67.661}.

\bibitem[{\citenamefont{Cr{\'e}peau et~al.}(2002)\citenamefont{Cr{\'e}peau,
  Gottesman, and Smith}}]{CGS02}
\bibinfo{author}{\bibfnamefont{C.}~\bibnamefont{Cr{\'e}peau}},
  \bibinfo{author}{\bibfnamefont{D.}~\bibnamefont{Gottesman}},
  \bibnamefont{and} \bibinfo{author}{\bibfnamefont{A.}~\bibnamefont{Smith}}, in
  \emph{\bibinfo{booktitle}{Proceedings of the Thiry-fourth Annual ACM
  Symposium on Theory of Computing}} (\bibinfo{publisher}{ACM},
  \bibinfo{address}{New York, NY, USA}, \bibinfo{year}{2002}), STOC '02, pp.
  \bibinfo{pages}{643--652}, ISBN \bibinfo{isbn}{1-58113-495-9},
  \urlprefix\url{http://doi.acm.org/10.1145/509907.510000}.

\bibitem[{\citenamefont{Broadbent et~al.}(2009)\citenamefont{Broadbent,
  Fitzsimons, and Kashefi}}]{BFK09}
\bibinfo{author}{\bibfnamefont{A.}~\bibnamefont{Broadbent}},
  \bibinfo{author}{\bibfnamefont{J.}~\bibnamefont{Fitzsimons}},
  \bibnamefont{and} \bibinfo{author}{\bibfnamefont{E.}~\bibnamefont{Kashefi}},
  in \emph{\bibinfo{booktitle}{Foundations of Computer Science, 2009. FOCS'09.
  50th Annual IEEE Symposium on}} (\bibinfo{organization}{IEEE},
  \bibinfo{year}{2009}), pp. \bibinfo{pages}{517--526}.

\bibitem[{\citenamefont{Morimae and Fujii}(2012)}]{morimae2012blind}
\bibinfo{author}{\bibfnamefont{T.}~\bibnamefont{Morimae}} \bibnamefont{and}
  \bibinfo{author}{\bibfnamefont{K.}~\bibnamefont{Fujii}},
  \bibinfo{journal}{Nature communications} \textbf{\bibinfo{volume}{3}},
  \bibinfo{pages}{1036} (\bibinfo{year}{2012}).

\bibitem[{\citenamefont{Barz et~al.}(2012)\citenamefont{Barz, Kashefi,
  Broadbent, Fitzsimons, Zeilinger, and Walther}}]{barz2012demonstration}
\bibinfo{author}{\bibfnamefont{S.}~\bibnamefont{Barz}},
  \bibinfo{author}{\bibfnamefont{E.}~\bibnamefont{Kashefi}},
  \bibinfo{author}{\bibfnamefont{A.}~\bibnamefont{Broadbent}},
  \bibinfo{author}{\bibfnamefont{J.~F.} \bibnamefont{Fitzsimons}},
  \bibinfo{author}{\bibfnamefont{A.}~\bibnamefont{Zeilinger}},
  \bibnamefont{and} \bibinfo{author}{\bibfnamefont{P.}~\bibnamefont{Walther}},
  \bibinfo{journal}{Science} \textbf{\bibinfo{volume}{335}},
  \bibinfo{pages}{303} (\bibinfo{year}{2012}).

\bibitem[{\citenamefont{Broadbent}(2015)}]{broadbent2015delegating}
\bibinfo{author}{\bibfnamefont{A.}~\bibnamefont{Broadbent}},
  \bibinfo{journal}{Canadian Journal of Physics} \textbf{\bibinfo{volume}{93}},
  \bibinfo{pages}{941} (\bibinfo{year}{2015}).

\bibitem[{\citenamefont{Aharonov et~al.}(2010)\citenamefont{Aharonov, Ben-Or,
  and Eban}}]{ABE}
\bibinfo{author}{\bibfnamefont{D.}~\bibnamefont{Aharonov}},
  \bibinfo{author}{\bibfnamefont{M.}~\bibnamefont{Ben-Or}}, \bibnamefont{and}
  \bibinfo{author}{\bibfnamefont{E.}~\bibnamefont{Eban}},
  \bibinfo{journal}{Proceedings of Innovations in Computer Science}
  (\bibinfo{year}{2010}).

\bibitem[{\citenamefont{Reichardt et~al.}(2013)\citenamefont{Reichardt, Unger,
  and Vazirani}}]{RUV}
\bibinfo{author}{\bibfnamefont{B.}~\bibnamefont{Reichardt}},
  \bibinfo{author}{\bibfnamefont{F.}~\bibnamefont{Unger}}, \bibnamefont{and}
  \bibinfo{author}{\bibfnamefont{U.}~\bibnamefont{Vazirani}},
  \bibinfo{journal}{Nature} \textbf{\bibinfo{volume}{496}},
  \bibinfo{pages}{7446} (\bibinfo{year}{2013}).

\bibitem[{\citenamefont{Fitzsimons and
  Kashefi}(2012)}]{fitzsimons2012unconditionally}
\bibinfo{author}{\bibfnamefont{J.~F.} \bibnamefont{Fitzsimons}}
  \bibnamefont{and} \bibinfo{author}{\bibfnamefont{E.}~\bibnamefont{Kashefi}},
  \bibinfo{journal}{arXiv preprint arXiv:1203.5217}  (\bibinfo{year}{2012}).

\bibitem[{\citenamefont{Morimae}(2014)}]{morimae2014verification}
\bibinfo{author}{\bibfnamefont{T.}~\bibnamefont{Morimae}},
  \bibinfo{journal}{Physical Review A} \textbf{\bibinfo{volume}{89}},
  \bibinfo{pages}{060302} (\bibinfo{year}{2014}).

\bibitem[{\citenamefont{Hayashi and Morimae}(2015)}]{hayashi2015verifiable}
\bibinfo{author}{\bibfnamefont{M.}~\bibnamefont{Hayashi}} \bibnamefont{and}
  \bibinfo{author}{\bibfnamefont{T.}~\bibnamefont{Morimae}},
  \bibinfo{journal}{Physical Review Letters} \textbf{\bibinfo{volume}{115}},
  \bibinfo{pages}{220502} (\bibinfo{year}{2015}).

\bibitem[{\citenamefont{Shamir}(1979)}]{Shamir1979}
\bibinfo{author}{\bibfnamefont{A.}~\bibnamefont{Shamir}},
  \bibinfo{journal}{Commun. ACM} \textbf{\bibinfo{volume}{22}},
  \bibinfo{pages}{612} (\bibinfo{year}{1979}), ISSN \bibinfo{issn}{0001-0782},
  \urlprefix\url{http://doi.acm.org/10.1145/359168.359176}.

\bibitem[{\citenamefont{Blakley}(1979)}]{Blakley1979}
\bibinfo{author}{\bibfnamefont{G.~R.} \bibnamefont{Blakley}},
  \bibinfo{journal}{Proc. of the National Computer Conference1979}
  \textbf{\bibinfo{volume}{48}}, \bibinfo{pages}{313} (\bibinfo{year}{1979}).

\bibitem[{\citenamefont{Hillery et~al.}(1999)\citenamefont{Hillery,
  Bu\ifmmode~\check{z}\else \v{z}\fi{}ek, and Berthiaume}}]{HBB99}
\bibinfo{author}{\bibfnamefont{M.}~\bibnamefont{Hillery}},
  \bibinfo{author}{\bibfnamefont{V.}~\bibnamefont{Bu\ifmmode~\check{z}\else
  \v{z}\fi{}ek}}, \bibnamefont{and}
  \bibinfo{author}{\bibfnamefont{A.}~\bibnamefont{Berthiaume}},
  \bibinfo{journal}{Phys. Rev. A} \textbf{\bibinfo{volume}{59}},
  \bibinfo{pages}{1829} (\bibinfo{year}{1999}),
  \urlprefix\url{http://link.aps.org/doi/10.1103/PhysRevA.59.1829}.

\bibitem[{\citenamefont{Cleve et~al.}(1999)\citenamefont{Cleve, Gottesman, and
  Lo}}]{CGL99}
\bibinfo{author}{\bibfnamefont{R.}~\bibnamefont{Cleve}},
  \bibinfo{author}{\bibfnamefont{D.}~\bibnamefont{Gottesman}},
  \bibnamefont{and} \bibinfo{author}{\bibfnamefont{H.-K.} \bibnamefont{Lo}},
  \bibinfo{journal}{Phys. Rev. Lett.} \textbf{\bibinfo{volume}{83}},
  \bibinfo{pages}{648} (\bibinfo{year}{1999}),
  \urlprefix\url{http://link.aps.org/doi/10.1103/PhysRevLett.83.648}.

\bibitem[{\citenamefont{Gottesman}(2000)}]{Got00}
\bibinfo{author}{\bibfnamefont{D.}~\bibnamefont{Gottesman}},
  \bibinfo{journal}{Phys. Rev. A} \textbf{\bibinfo{volume}{61}},
  \bibinfo{pages}{042311} (\bibinfo{year}{2000}),
  \urlprefix\url{http://link.aps.org/doi/10.1103/PhysRevA.61.042311}.

\bibitem[{\citenamefont{Zhang et~al.}(2005)\citenamefont{Zhang, Li, and
  Man}}]{zhang2005multiparty}
\bibinfo{author}{\bibfnamefont{Z.-j.} \bibnamefont{Zhang}},
  \bibinfo{author}{\bibfnamefont{Y.}~\bibnamefont{Li}}, \bibnamefont{and}
  \bibinfo{author}{\bibfnamefont{Z.-X.} \bibnamefont{Man}},
  \bibinfo{journal}{Physical Review A} \textbf{\bibinfo{volume}{71}},
  \bibinfo{pages}{044301} (\bibinfo{year}{2005}).

\bibitem[{\citenamefont{Markham and Sanders}(2008)}]{DSa08}
\bibinfo{author}{\bibfnamefont{D.}~\bibnamefont{Markham}} \bibnamefont{and}
  \bibinfo{author}{\bibfnamefont{B.~C.} \bibnamefont{Sanders}},
  \bibinfo{journal}{Physical Review A} \textbf{\bibinfo{volume}{78}},
  \bibinfo{pages}{042309} (\bibinfo{year}{2008}).

\bibitem[{\citenamefont{Rohde et~al.}(2012)\citenamefont{Rohde, Fitzsimons, and
  Gilchrist}}]{Rohde}
\bibinfo{author}{\bibfnamefont{P.~P.} \bibnamefont{Rohde}},
  \bibinfo{author}{\bibfnamefont{J.~F.} \bibnamefont{Fitzsimons}},
  \bibnamefont{and}
  \bibinfo{author}{\bibfnamefont{A.}~\bibnamefont{Gilchrist}},
  \bibinfo{journal}{Phys. Rev. Lett.} \textbf{\bibinfo{volume}{109}},
  \bibinfo{pages}{150501} (\bibinfo{year}{2012}),
  \urlprefix\url{http://link.aps.org/doi/10.1103/PhysRevLett.109.150501}.

\bibitem[{\citenamefont{Tan et~al.}(2016)\citenamefont{Tan, Kettlewell, Ouyang,
  Chen, and Fitzsimons}}]{tan2016quantum}
\bibinfo{author}{\bibfnamefont{S.-H.} \bibnamefont{Tan}},
  \bibinfo{author}{\bibfnamefont{J.~A.} \bibnamefont{Kettlewell}},
  \bibinfo{author}{\bibfnamefont{Y.}~\bibnamefont{Ouyang}},
  \bibinfo{author}{\bibfnamefont{L.}~\bibnamefont{Chen}}, \bibnamefont{and}
  \bibinfo{author}{\bibfnamefont{J.~F.} \bibnamefont{Fitzsimons}},
  \bibinfo{journal}{Scientific Reports} \textbf{\bibinfo{volume}{6}},
  \bibinfo{pages}{33467} (\bibinfo{year}{2016}).

\bibitem[{\citenamefont{Ouyang et~al.}(2015)\citenamefont{Ouyang, Tan, and
  Fitzsimons}}]{ouyang2015quantum}
\bibinfo{author}{\bibfnamefont{Y.}~\bibnamefont{Ouyang}},
  \bibinfo{author}{\bibfnamefont{S.-H.} \bibnamefont{Tan}}, \bibnamefont{and}
  \bibinfo{author}{\bibfnamefont{J.}~\bibnamefont{Fitzsimons}},
  \bibinfo{journal}{arXiv preprint arXiv:1508.00938}  (\bibinfo{year}{2015}).

\bibitem[{\citenamefont{Broadbent and Jeffery}(2015)}]{BJe15}
\bibinfo{author}{\bibfnamefont{A.}~\bibnamefont{Broadbent}} \bibnamefont{and}
  \bibinfo{author}{\bibfnamefont{S.}~\bibnamefont{Jeffery}}, in
  \emph{\bibinfo{booktitle}{Annual Cryptology Conference}}
  (\bibinfo{organization}{Springer}, \bibinfo{year}{2015}), pp.
  \bibinfo{pages}{609--629}.

\bibitem[{\citenamefont{Dulek et~al.}(2016)\citenamefont{Dulek, Schaffner, and
  Speelman}}]{DSS16}
\bibinfo{author}{\bibfnamefont{Y.}~\bibnamefont{Dulek}},
  \bibinfo{author}{\bibfnamefont{C.}~\bibnamefont{Schaffner}},
  \bibnamefont{and} \bibinfo{author}{\bibfnamefont{F.}~\bibnamefont{Speelman}},
  pp. \bibinfo{pages}{3--32} (\bibinfo{year}{2016}).

\bibitem[{\citenamefont{Yu et~al.}(2014)\citenamefont{Yu,
  {P{{\'e}}rez-{D}elgado}, and Fitzsimons}}]{YPF14}
\bibinfo{author}{\bibfnamefont{L.}~\bibnamefont{Yu}},
  \bibinfo{author}{\bibfnamefont{C.~A.} \bibnamefont{{P{{\'e}}rez-{D}elgado}}},
  \bibnamefont{and} \bibinfo{author}{\bibfnamefont{J.~F.}
  \bibnamefont{Fitzsimons}}, \bibinfo{journal}{Phys. Rev. A}
  \textbf{\bibinfo{volume}{90}}, \bibinfo{pages}{050303}
  (\bibinfo{year}{2014}),
  \urlprefix\url{http://link.aps.org/doi/10.1103/PhysRevA.90.050303}.

\bibitem[{\citenamefont{Newman and Shi}(2016)}]{NewmanShi2016}
\bibinfo{author}{\bibfnamefont{M.}~\bibnamefont{Newman}} \bibnamefont{and}
  \bibinfo{author}{\bibfnamefont{Y.}~\bibnamefont{Shi}} (\bibinfo{year}{2016}),
  \bibinfo{note}{private communication}.

\bibitem[{\citenamefont{Zhou et~al.}(2000)\citenamefont{Zhou, Leung, and
  Chuang}}]{ZLC00}
\bibinfo{author}{\bibfnamefont{X.}~\bibnamefont{Zhou}},
  \bibinfo{author}{\bibfnamefont{D.~W.} \bibnamefont{Leung}}, \bibnamefont{and}
  \bibinfo{author}{\bibfnamefont{I.~L.} \bibnamefont{Chuang}},
  \bibinfo{journal}{Phys. Rev. A} \textbf{\bibinfo{volume}{62}},
  \bibinfo{pages}{052316} (\bibinfo{year}{2000}),
  \urlprefix\url{http://link.aps.org/doi/10.1103/PhysRevA.62.052316}.

\bibitem[{\citenamefont{Nielsen and Chuang}(2000)}]{nielsen-chuang}
\bibinfo{author}{\bibfnamefont{M.~A.} \bibnamefont{Nielsen}} \bibnamefont{and}
  \bibinfo{author}{\bibfnamefont{I.~L.} \bibnamefont{Chuang}},
  \emph{\bibinfo{title}{{Quantum Computation and Quantum Information}}}
  (\bibinfo{publisher}{Cambridge University Press}, \bibinfo{year}{2000}),
  \bibinfo{edition}{2nd} ed.

\bibitem[{\citenamefont{Dunjko et~al.}(2014)\citenamefont{Dunjko, Fitzsimons,
  Portmann, and Renner}}]{dunjko2014composable}
\bibinfo{author}{\bibfnamefont{V.}~\bibnamefont{Dunjko}},
  \bibinfo{author}{\bibfnamefont{J.~F.} \bibnamefont{Fitzsimons}},
  \bibinfo{author}{\bibfnamefont{C.}~\bibnamefont{Portmann}}, \bibnamefont{and}
  \bibinfo{author}{\bibfnamefont{R.}~\bibnamefont{Renner}}, in
  \emph{\bibinfo{booktitle}{Advances in Cryptology--ASIACRYPT 2014}}
  (\bibinfo{publisher}{Springer}, \bibinfo{year}{2014}), pp.
  \bibinfo{pages}{406--425}.

\end{thebibliography}

\end{document}